%% file: main.tex
\documentclass[10pt,twocolumn,letterpaper]{article}

\usepackage[pagenumbers]{iccv} 


\definecolor{iccvblue}{rgb}{0.21,0.49,0.74}
\usepackage[pagebackref,breaklinks,colorlinks,allcolors=iccvblue]{hyperref}

\def\paperID{8786} 
\def\confName{ICCV}
\def\confYear{2025}

\title{DeepSound-V1: Start to Think Step-by-Step in the Audio Generation from Videos}

\author{
    \textit{Yunming Liang$^1$$^*$, Zihao Chen$^1$$^*$, Chaofan Ding$^1$, Xinhan Di$^1$} \\
    $^1$AI Lab, Giant Network.\\
    \small\texttt{\{liangyunming, chenzihao, dingchaofan, dixinhan\}@ztgame.com}
}

\usepackage[absolute,overlay]{textpos}  

\begin{document}
\maketitle
\input{sec/0_abstract}

\input{sec/1_intro}

\input{sec/2_related_work}
\input{sec/3_method}

\input{sec/4_experiment}

\input{sec/5_discussion}
{
    \small
    \bibliographystyle{ieeenat_fullname}
    \bibliography{main}
}

\end{document}

%% file: sec/0_abstract.tex
\begin{abstract}
Currently, high-quality, synchronized audio is synthesized from video and optional text inputs using various multi-modal joint learning frameworks. However, the precise alignment between the visual and generated audio domains remains far from satisfactory. One key factor is the lack of sufficient temporal and semantic alignment annotations in open-source video-audio and text-audio benchmarks. Therefore, we propose a framework for audio generation from videos, leveraging the internal chain-of-thought (CoT) of a multi-modal large language model (MLLM) to enable step-by-step reasoning without requiring additional annotations. Additionally, a corresponding multi-modal reasoning dataset is constructed to facilitate the learning of initial reasoning in audio generation. In the experiments, we demonstrate the effectiveness of the proposed framework in reducing misalignment (voice-over) in generated audio and achieving competitive performance compared to various state-of-the-art models. The evaluation results show that the proposed method outperforms state-of-the-art approaches across multiple metrics. Specifically, the $FD_{PaSST}$ indicator is reduced by up to 10.07\%, the $FD_{PANNs}$ indicator by up to 11.62\%, and the $FD_{VGG}$ indicator by up to 38.61\%. Furthermore, the IS indicator improves by up to 4.95\%, the IB-score indicator increases by up to 6.39\%, and the DeSync indicator is reduced by up to 0.89\%. 
\end{abstract}

%% file: sec/1_intro.tex
\section{Introduction}

Foley referring to the process of generating high-quality, synchronized sound effects to enhance video content, should satisfy two critical objectives: semantic coherence and temporal precision. This task demands models to interpret scene semantics and audio-visual relationships while maintaining precise alignment ~\cite{cheng2024taming, luo2023diff, zhang2024foleycrafter, SpecVQGAN_Iashin_2021, wang2024frieren}. Current methodologies fall into two primary categories: 1) Architecture-specialized approaches that optimize alignment through task-specific modules ~\cite{zhang2024foleycrafter, wang2024v2a, li2024tri}, and 2) Unified architecture strategies leveraging diffusion-based transformer (DiT) frameworks to jointly model visual-audio dependencies ~\cite{cheng2024taming,chen2024video}.  

A variety of Video-to-Audio (V2A) models~\cite{cheng2024taming, viertola2024temporally} are trained on audio-visual datasets comprising noisy, in-the-wild samples, such as VGGSound~\cite{chen2020vggsound} or AudioSet~\cite{gemmeke2017audio}, where the audio-visual relevance is not explicitly ensured. In V2A generation, a critical challenge arises when the audio contains voice-over unrelated to the visual content, this mismatch introduces significant noise during the generation process. 

\begin{figure}[t]
    \centering
    \includegraphics[width=\columnwidth]{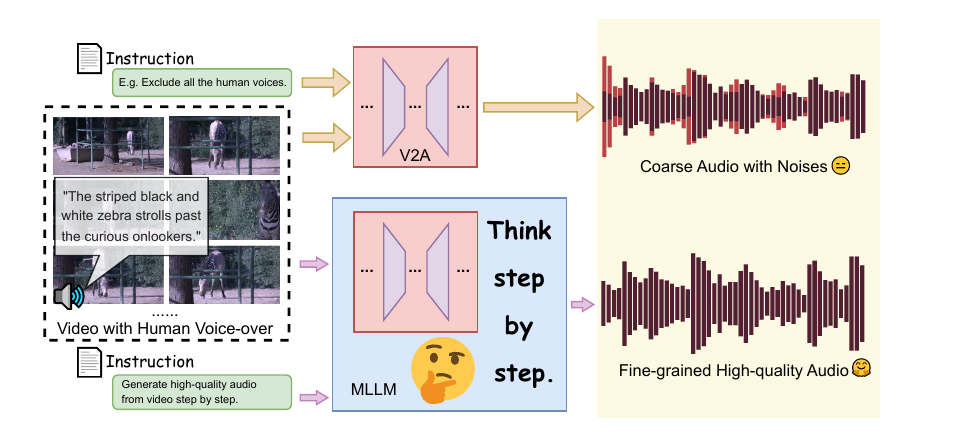}
    \caption{Current V2A models \cite{zhang2024foleycrafter, cheng2024taming, chen2024yingsound} (upper) represent existing approaches. The proposed DeepSound (below) are designed to follow a step-by-step reasoning process to eliminate voice-over.}
    \label{fig:fig1}
\end{figure}


Inspired via rapid development of step-by-step large language reasoning models, exemplified by o1~\cite{openai2024o1}, o1-mini~\cite{openai2024o1mini}, o3-mini~\cite{openai2024o3mini}, grok3~\cite{grok3}, claude3.7~\cite{claude37}, and R1~\cite{deepseekai2025deepseekr1incentivizingreasoningcapability}, and promising reasoning ability of Multi-modal large language model (MLLM)~\cite{xu2024llava, yan2025position, du2025virgo}, we propose a novel MLLM-based framework that explicitly incorporates structured reasoning mechanisms for video-guided audio generation, specifically addressing the voice-over that is misaligned with the visual content. Our approach aims to release the audio-visual misalignment (e.g., alleviate off-screen narration and temporal inconsistencies) through cross-modal causal reasoning. This framework enables audio generation from videos as an initial step-by-step process, guided by the internal chain-of-thoughts (CoTs) of MLLM, without requiring additional annotations. Additionally, a corresponding multi-modal reasoning dataset is constructed to facilitate the learning of initial thinking in audio generation. In the experiment, our proposed framework demonstrates improved performance in comparison with the state-of-the-art models.

%% file: sec/2_related_work.tex
\section{related work}
\subsection{Multi-modal Alignment and Condition} 

Recent advancements in V2A synthesis focus on establishing semantic and temporal coherence through diverse training strategies. While foundational methods ~\cite{chen2020generating, du2023conditional, iashin2021taming, mei2024foleygen, pascual2024masked, su2023physics, wang2024v2a, yang2024draw} employ direct audio-visual pair supervision with generative objectives, newer approaches explore hierarchical alignment paradigms. Certain studies~\cite{zhang2024foleycrafter, huang2024rhythmic, chung2024t, lee2024video} adopt a two-stage process ~\cite{huang2024rhythmic, lee2024video, mo2024text, zhang2024foleycrafter} and one-stage process ~\cite{cheng2024taming,chen2024yingsound} for high quality audio generation. However, the presence of noise in the data inevitably leads to audio containing unintended voice-over. To address this, we propose the DeepSound framework, which generates audio from video step-by-step, detects voice-over, and refines coarse audio into fine-grained audio.


\subsection{Multimodal Generation} 
Multimodal generation models generate samples composed of multiple modalities (e.g., video and audio at the same time, text and speech~\cite{xie2024mini, fu2024vita, zeng2024glm, liu2025ola} at the same time). While multimodal generation presents inherently greater complexity, state-of-the-art methodologies~\cite{kim2024versatile, ruan2023mm, tang2024codi, tang2023any} still fall short of matching the performance of specialized video-to-audio systems. To optimize V2A synthesis and minimize the generation of unwanted voice-over artifacts, we propose a framework that combines MLLM with CoT reasoning, incorporating step-by-step guidance throughout the process.

\subsection{Multimodal Reasoning} 
Visual reasoning requires models to integrate visual perception with high-level cognition~\cite{johnson2017clevr, malkinski2023review}. Standard evaluation tasks include Visual Question Answering (VQA)~\cite{ishmam2024image, li2022joint} and Visual Entailment~\cite{song2022clip}, which assess multimodal consistency. Traditional vision-language models employ neural symbolic approaches~\cite{amizadeh2020neuro, choi2024towards} to explicitly structure reasoning processes. Modern methods leverage large language models (LLMs) to interpret visual tasks~\cite{liu2023visual, yu2024evagaussians}, enhanced by optimized visual encoding strategies~\cite{jin2024chat, li2024tokenpacker, liu2023visual} that generate cognition-focused tokens. Techniques like prompt tuning~\cite{zamfirescu2023johnny}, in-context learning, and supervised fine-tuning~\cite{shen2024rethinking} further augment visual reasoning. Inspired via rapid develop of step-by-step large language reasoning models, exemplified by o1~\cite{openai2024o1}, o1-mini~\cite{openai2024o1mini}, o3-mini~\cite{openai2024o3mini}, grok3~\cite{grok3}, claude3.7~\cite{claude37}, and R1~\cite{deepseekai2025deepseekr1incentivizingreasoningcapability}. LLaVA-CoT~\cite{xu2024llava}, a vision-language model designed for systematic reasoning, achieves inference-time scalability through stage-level beam search. However, the critical issue of voice-over artifacts in V2A synthesis has yet to be extensively explored. To overcome this limitation, we propose the DeepSound framework, a step-by-step CoT architecture that explicitly integrates cross-modal reasoning to mitigate voice-over artifacts in video-guided audio generation.

%% file: sec/3_method.tex
\begin{figure*}[t]
    \centering    \includegraphics[width=1\textwidth]{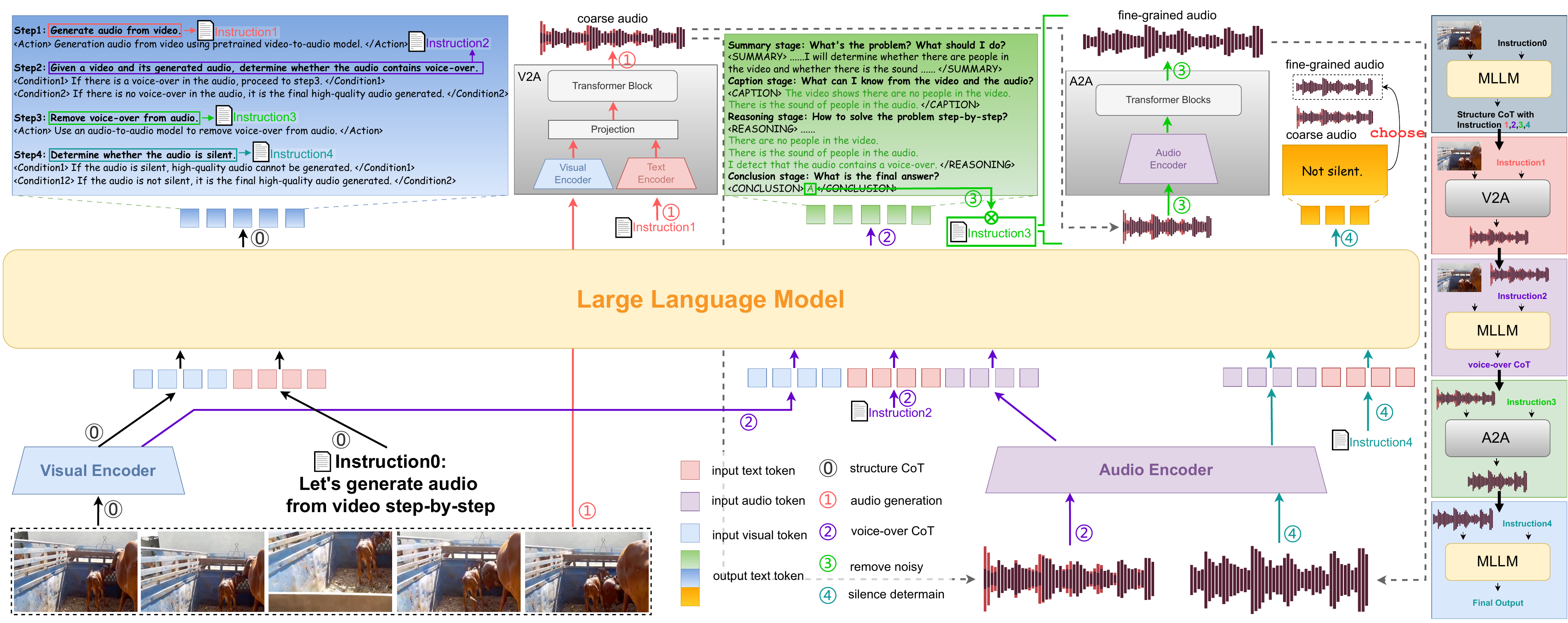}
    \caption{\textbf{Overview of DeepSound.} The model employs a step-by-step reasoning process to generate audio from video. In the first step, it generates a coarse audio from the input video. The second step identifies voice-over components by analyzing both the coarse audio and the video. The third step removes the detected voice-over elements from the audio. Finally, the model determines whether the resulting audio is silent or not.}
    \label{fig:fig2}
\end{figure*}

\section{method}
\subsection{Overview}
A DeepSound (Figure~\ref{fig:fig2}) framework is proposed to facilitate a progressive, step-by-step reasoning process that both enhances the reasoning ability in the generation of audio and the quality of generated audio. Through multi-modal and internal CoT generation to guild the generation of audio from video, the proposed framework is expected to release the side-effect of noise (voice-over) in the generated audio aiming to improve the quality of the audio.
The framework is composed with three modules, a module for generating audio from video $\text{M}_{\text{Audio}}$, a multi-modal reasoning module $\text{M}_{\text{Reasoning}}$ and an audio editing module $\text{M}_{\text{Edit}}$. The four generated reasoning steps are represented as the following:

\begin{itemize}
    \item \textbf{Step 1:} Generate audio from video.
    
    \item \textbf{Step 2:} Given a video and its generated audio, determine whether the audio contains voice-over.
    
    \item \textbf{Step 3:} Remove voice-over from audio.

    \item \textbf{Step 4:} Determine whether the audio is silent.
\end{itemize}
The above process of internal generated CoT in the process of the audio generation is represented as the following:

\begin{equation}
\begin{aligned}
    \text{M}_{\text{Reasoning}} &: (X, V) \mapsto \text{CoT}_{\text{structure}}, \\
    \text{M}_{\text{audio}} &: (X, V, \text{CoT}_{\text{structure}}) \mapsto \text{Audio}_{\text{coarse}}, \\
    \text{M}_{\text{Reasoning}} &: (\text{Audio}_{\text{coarse}}, V, \text{CoT}_{\text{structure}}) \mapsto \text{CoT}_{\text{detail}}, \\
    \text{M}_{\text{Edit}} &: (\text{Audio}_{\text{coarse}}, \text{CoT}_{\text{structure}}, \text{CoT}_{\text{detail}}) 
    \mapsto \text{Audio}_{\text{FG}}.
\end{aligned}
\end{equation}
where $X$, $V$ denote the provided textual instruction and input video clip, respectively. $\text{M}_{\text{Audio}}$, $\text{M}_{\text{Resoning}}$, and $\text{M}_{\text{Edit}}$ are three core components comprising the DeepSound framework. $\text{CoT}_{\text{structure}}$ serves as the outer-layer cognitive architecture governing hierarchical reasoning processes, while $\text{Audio}_{\text{coarse}}$ functions as the core synthesis module responsible for generating preliminary audio outputs through visual-temporal grounding. $\text{Audio}_{\text{FG}}$ denotes the fine-grained audio output generated by our CoT-guided reasoning framework, achieving near-complete suppression of voice-over artifacts through iterative cross-modal refinement.

\subsection{Multiple Multi-Modal Modules}

\subsubsection{Video-Audio Generation Module}
To optimize the coarse audio generation process, we define the corresponding objective function:

\begin{equation}
\begin{aligned}
    \text{M}_{\text{audio}} &: (X, V, \text{CoT}_{\text{structure}}) \mapsto \text{Audio}_{\text{coarse}}, \\
    \min_{\theta_{\text{audio}}} \mathbb{E}_{(X, V) \sim \mathcal{D}} & \Big[ \mathcal{L}_{\text{audio\_gen}}(\text{M}_{\text{audio}}(X, V, \text{CoT}_{\text{structure}}), \text{Audio}_{\text{gt}}) \Big], \\
    \mathcal{L}_{\text{audio\_gen}} &= MSE \bigl( \text{Audio}_{\text{coarse}}, \text{Audio}_{\text{gt}} \bigr)
\end{aligned}
\end{equation}

\begin{figure*}[t]
    \centering    \includegraphics[width=1\textwidth]{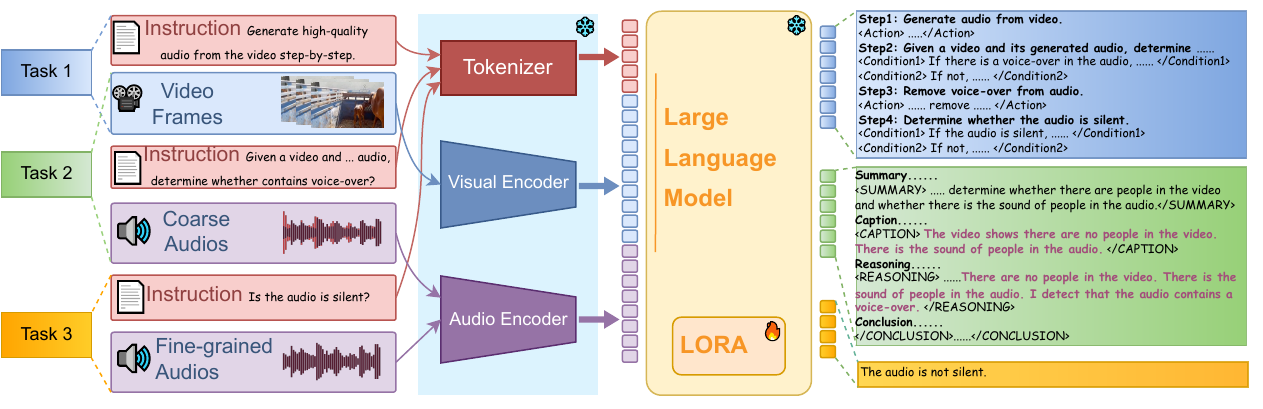}
    \caption{\textbf{Overview of Dual Multi-Modal Reasoning Learning.} $CoT_{structure}$ represents the internal reasoning steps within the overall audio generation process. $CoT_{detail}$ refers to the step-by-step procedure for identifying voice-over components from the coarse audio and video.}
    \label{fig:fig3}
\end{figure*}

\subsubsection{Multimodal Large Language Model}

where leveraging the structured reasoning framework \(\text{CoT}_{\text{structure}}\), the multimodal audio generation module \(\text{M}_{\text{audio}}\) generates coarse audio \(\text{Audio}_{\text{coarse}}\) conditioned on the input video clip \(V\) and textual instructions \(X\). The optimization objective aims to learn the parameters \(\theta_{\text{audio}}\) by minimizing the mean square error (MSE) loss \(\mathcal{L}_{\text{audio\_gen}}\) between the generated audio and the ground truth audio \(\text{Audio}_{\text{gt}}\), ensuring that the synthesized audio closely aligns with the reference audio.

\begin{equation}
\begin{aligned}
    \text{M}_{\text{Reasoning}} &: \Bigl( \text{Audio}_{\text{coarse}}, V, \text{CoT}_{\text{structure}} \Bigr) \mapsto \text{CoT}_{\text{detail}}, \\
    \min_{\theta_{\text{CoT}_{\text{detail}}}} &\mathbb{E}_{(X, V, \text{Audio}_{\text{coarse}}) \sim \mathcal{D}} \Big[ \mathcal{L}_{\text{CoT}_{\text{detail}}} \Big( \\
    &\quad f^{\text{CoT}_{\text{detail}}}_{\theta}(X, V, \text{Audio}_{\text{coarse}}), Y_{\text{CoT}_{\text{detail}}} \Big) \Big], \\
    \mathcal{L}_{\text{CoT}_{\text{detail}}} &= \mathcal{L}_{\text{detail\_format}} + \mathcal{L}_{\text{detail\_keyword}}
\end{aligned}   
\end{equation}

Building upon a MLLM that integrates visual, audio, and textual modalities, we fine-tune this MLLM using our pre-constructed CoT dataset. Subsequently, we input the first-stage generated \(\text{Audio}_{\text{coarse}}\), along with the corresponding video \(V\) and textual instructions \(X\), into the fine-tuned \(\text{M}_{\text{Reasoning}}\) module to determine whether the video-guided audio generation contains voice-over artifacts.

\subsubsection{Audio Editing Module}

To optimize the detailed reasoning process, we aim to learn the parameters \(\theta_{\text{CoT}_{\text{detail}}}\) by minimizing the composite CoT detail loss \(\mathcal{L}_{\text{CoT}_{\text{detail}}}\). This loss function consists of two complementary components: \(\mathcal{L}_{\text{detail\_format}}\), which maintains structural integrity, and \(\mathcal{L}_{\text{detail\_keyword}}\), which preserves content granularity. The optimization objective ensures that the generated CoT details align with the ground truth \(Y_{\text{CoT}_{\text{detail}}}\), thereby enhancing the coherence and informativeness of the reasoning outputs.

    

\begin{equation}
\begin{aligned}
    \text{M}_{\text{Edit}} :\  (\text{Audio}_{\text{coarse}},\, \text{CoT}_{\text{structure}},\, \text{CoT}_{\text{detail}}) 
    \mapsto \text{Audio}_{\text{FG}}, \\
    \min_{\theta_{\text{audio}}}\  \mathbb{E}_{(\text{Audio}_{\text{coarse}},\, \text{CoT}_{\text{structure}},\, \text{CoT}_{\text{detail}}) \sim \mathcal{D}}  
    \Big[ \mathcal{L}_{\text{audio\_remove}} \Big(   
    \text{M}_{\text{Edit}} \\ (\text{Audio}_{\text{coarse}},\, \text{CoT}_{\text{structure}},\, \text{CoT}_{\text{detail}}),\, \text{Audio}_{\text{gt}} \Big) \Big],
\end{aligned}
\end{equation}

\begin{align}
    \mathcal{L}_{\text{audio\_remove}} =\ & \lVert a - \hat{a} \rVert + \sum_{s=0}^{S-1} 
    \lVert \mathbf{A}^{(s)} - \hat{\mathbf{A}}^{(s)} \rVert.
\end{align}

To further refine the audio generation, \(\text{M}_{\text{Edit}}\) edits the coarse audio \(\text{Audio}_{\text{coarse}}\) following the structured reasoning outputs \(\text{CoT}_{\text{structure}}\) and \(\text{CoT}_{\text{detail}}\), ultimately producing the fine-grained audio \(\text{Audio}_{\text{FG}}\). The optimization process aims to learn the parameters \(\theta_{\text{audio}}\) by minimizing the voice-over removal loss \(\mathcal{L}_{\text{audio\_remove}}\), ensuring that the edited audio closely aligns with the ground truth \(\text{Audio}_{\text{gt}}\).


Furthermore, our framework incorporates a conditional fallback mechanism during post-processing: after voice-over removal, audio clips undergo silence detection via an MLLM. If the MLLM determines that a clip is silent, the system reverts to the original pre-removal audio for that segment.

The loss function \(\mathcal{L}_{\text{audio\_remove}}\) is designed to guide the optimization by balancing time-domain and frequency-domain consistency. Specifically, \(\mathbf{A}^{(s)}\) and \(\hat{\mathbf{A}}^{(s)}\) represent the audio content in the frequency domain, while \(a\) and \(\hat{a}\) correspond to the transformed information in the time domain. The first term \(\lVert a - \hat{a} \rVert\) measures the reconstruction error in the time domain, ensuring temporal coherence. The second term \(\sum_{s=0}^{S-1} \lVert \mathbf{A}^{(s)} - \hat{\mathbf{A}}^{(s)} \rVert\) evaluates the discrepancy in the frequency domain across different subbands, preserving spectral fidelity. By jointly minimizing these terms, the model effectively refines the audio while mitigating excessive content removal.

\subsection{Dual Multi-Modal Reasoning Learning}
\begin{equation}
\begin{aligned}
    \text{M}_{\text{Reasoning}} &: (X, V) \mapsto
    \text{CoT}_{\text{structure}} , \\
       \text{M}_{\text{Reasoning}} &:  \Bigl( \text{Audio}_{\text{coarse}},\; 
            V,\; 
            \text{CoT}_{\text{structure}} \Bigr) \mapsto \text{CoT}_{\text{detail}} \\
\end{aligned}
\end{equation}

To achieve internal reasoning steps in the generation process of audio through the proposed framework, the reasoning module first generates step-by-step  $\text{CoT}_{\text{structure}}$~\ref{fig:fig4} given an instruction $X$ and a video $V$. 

Secondly, all three modules $\text{M}_{\text{Audio}}$, $\text{M}_{\text{Reasoning}}$, and $\text{M}_{\text{Edit}}$ operate under the generated guidance. Specifically, $\text{M}_{\text{Audio}}$ produces coarse audio $\text{Audio}_{\text{coarse}}$ given video $V$ and $\text{CoT}_{\text{structure}}$. $\text{M}_{\text{Reasoning}}$ generates additional step-by-step $\text{CoT}_{\text{detail}}$ given $\text{Audio}_{\text{coarse}}$, $V$, and $\text{CoT}_{\text{structure}}$. The subsequent processing steps dynamically adjust based on the intermediate reasoning results from $\text{CoT}_{\text{detail}}$.

To optimize the reasoning process, we define the following objective function:
\begin{multline}
    \min_{\theta_{\text{CoT}}, \theta_{\text{CoT}_{\text{detail}}}} \mathbb{E}_{(X, V) \sim \mathcal{D}} \Big[ \mathcal{L}_{\text{CoT}}(f^{\text{CoT}}_{\theta}(X, V), Y_{\text{CoT}}) \Big] \\
    + \mathbb{E}_{(X, V, \text{Audio}_{\text{coarse}}) \sim \mathcal{D}} \Big[ \mathcal{L}_{\text{CoT}_{\text{detail}}}(f^{\text{CoT}_{\text{detail}}}_{\theta}(X, V, \text{Audio}_{\text{coarse}}), \\
    Y_{\text{CoT}_{\text{detail}}}) \Big]
\end{multline}
where $Y_{\text{CoT}}$ and $Y_{\text{CoT}_{\text{detail}}}$ represent the ground truth reasoning sequences for $\text{CoT}_{\text{structure}}$ and $\text{CoT}_{\text{detail}}$, respectively. The loss functions $\mathcal{L}_{\text{CoT}}$ and $\mathcal{L}_{\text{CoT}_{\text{detail}}}$ measure the deviation between the generated reasoning steps and the optimal sequences. The first term ensures structured step-by-step reasoning from input $(X, V)$, while the second term refines the reasoning based on intermediate audio features, enhancing the coherence and precision of the final generated output.

In the proposed DeepSound framework, joint training is implemented to optimize two components: $format$ to enforce structural constraints on output format and $keyword$ to enhance semantic accuracy in keyword extraction. These loss terms are derived through multimodal reasoning from synchronized visual and audio inputs. The composite objective function is defined as follows:
\begin{equation}
\begin{aligned}
    \mathcal{L}_{\text{MLLM}} &= \mathcal{L}_{\text{CoT}} + \mathcal{L}_{\text{CoT}_{\text{detail}}} \\
    &= \mathcal{L}_{\text{format}} + \mathcal{L}_{\text{keyword}}, \\
    \mathcal{L}_{\text{format}} &= \mathcal{L}_{\text{format}}^{1} + \mathcal{L}_{\text{format\_SM}}^{2} + \mathcal{L}_{\text{format\_CP}}^{2} \\
    &\quad + \mathcal{L}_{\text{format\_RN}}^{2} + \mathcal{L}_{\text{format\_CC}}^{2}, \\
    \mathcal{L}_{\text{keyword}} &= \mathcal{L}_{\text{keyword\_CP}} + \mathcal{L}_{\text{keyword\_RN}} + \mathcal{L}_{\text{keyword\_CC}}
\end{aligned}
\end{equation}
where $\mathcal{L}_{\text{MLLM}}$ represents the total loss of MLLM derived from step-by-step reasoning. Specifically, $\mathcal{L}_{\text{CoT}}$ corresponds to the loss associated with generating structured reasoning steps, ensuring coherence in the reasoning process. Meanwhile, $\mathcal{L}_{\text{CoT}_{\text{detail}}}$ refines the reasoning by incorporating additional details derived from intermediate representations, improving the overall consistency and accuracy of the reasoning output. Moreover, $\mathcal{L}_{\text{format}}$ represents the overall format loss, $\mathcal{L}_{\text{format}}^{1}$ corresponds to $\text{CoT}_{\text{structure}}$ loss, and $\mathcal{L}_{\text{format\_SM}}^{2}$, $\mathcal{L}_{\text{format\_CP}}^{2}$, $\mathcal{L}_{\text{format\_RN}}^{2}$, and $\mathcal{L}_{\text{format\_CC}}^{2}$ correspond to losses at the summary, caption, reasoning, and conclusion stages respectively. $\mathcal{L}_{\text{keyword}}$ is designed to enhance voice-over detection accuracy in audiovisual inputs by aligning cross-modal features. $\mathcal{L}_{\text{keyword\_CP}}$, $\mathcal{L}_{\text{keyword\_RN}}$, and $\mathcal{L}_{\text{keyword\_CC}}$ correspond to keyword extraction losses at different reasoning stages.

%% file: sec/4_experiment.tex
\section{Experiments}
\subsection{Dataset}
\subsubsection{Open-sourced Dataset}
VGGSound~\cite{chen2020vggsound} is a large-scale audio-visual dataset designed for audio recognition and multimodal learning tasks. It consists of over $200,000$ video clips sourced from YouTube, covering 309 diverse audio classes. The dataset is divided into training and testing sets, with a total duration of more than 550 hours. 

\subsubsection{Custom Dataset}
We build an 18k multimodal CoT V2A dataset based on VGGSound to generate high-quality audio from video. Based on CoT reasoning and CoT-like guidance \cite{xu2025llavacotletvisionlanguage}, we utilize a professional annotation team to label the dataset. We develop a CoT reasoning framework to guide subsequent generation of high-quality audio from video tasks, as illustrated in Figure~\ref{fig:fig2}. Specifically, a step-by-step instruction process with video and audio input is designed to enable efficient and accurate voice-over judgement. As shown in Figure~\ref{fig:fig4}, \texttt{<SUMMARY>}\texttt{</SUMMARY>} effective decomposition of human and human voices for the task of judging voice-over, while \texttt{<CAPTION>}\texttt{</CAPTION>} describes the people in the video and the voices of the people in the audio. During the \texttt{<REASONING>}\texttt{</REASONING>} stage, the reasoning process is divided into four steps: Step 1. determine that the judgment of the voice-over is based on a rule. Step 2. specify the rule of whether to include the voice-over based on the situation of the person and the voice. Step 3. judge whether there is a person in the video and whether there is a human voice in the audio. Step 4. conclusion and give the answer. Each stage is initiated at the model’s discretion, without external prompt engineering frameworks or additional prompting. Specifically, we provide the model with four pairs of special tags: \texttt{<SUMMARY>}\texttt{</SUMMARY>}, \texttt{<CAPTION>}\texttt{</CAPTION>}, \texttt{<REASONING>}\texttt{</REASONING>}, and \texttt{<CONCLUSION>} \texttt{</CONCLUSION>}. These tags correspond to summarizing the response approach, describing relevant image and audio content, conducting reasoning, and preparing a final answer, respectively. Detailed statistics and the construction process are illustrated in Figure~\ref{fig:fig4}. Additionally, we constructed 1.8k $\text{CoT}_{\text{structure}}$ samples for fine-tuning.



\begin{figure}[t]
    \centering
    \includegraphics[width=\columnwidth]{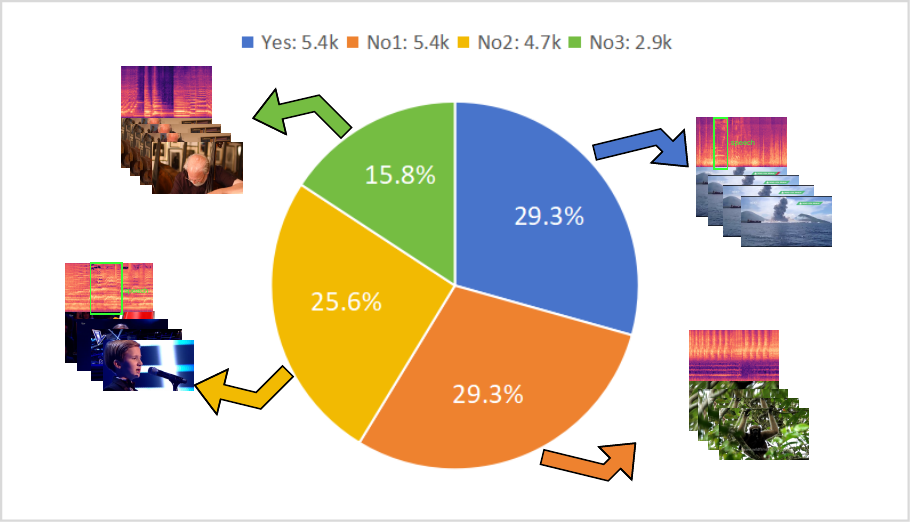}
    \caption{The voice-over labels are divided into four categories based on the presence or absence of people and human voices. The label \textbf{"Yes"} indicates that the sample contains voice-over, and the labels \textbf{"No1"}, \textbf{"No2"}, and \textbf{"No3"} indicate that the sample does not contain voice-over. Specifically, \textbf{"No1"} means the video contains neither people nor human voices, \textbf{"No2"} means the video contains both people and human voices, and \textbf{"No3"} means the video contains people but without human voices.}
    \label{fig:fig1}
\end{figure}

\begin{figure*}[t]
    \centering    \includegraphics[width=1.25\textwidth]{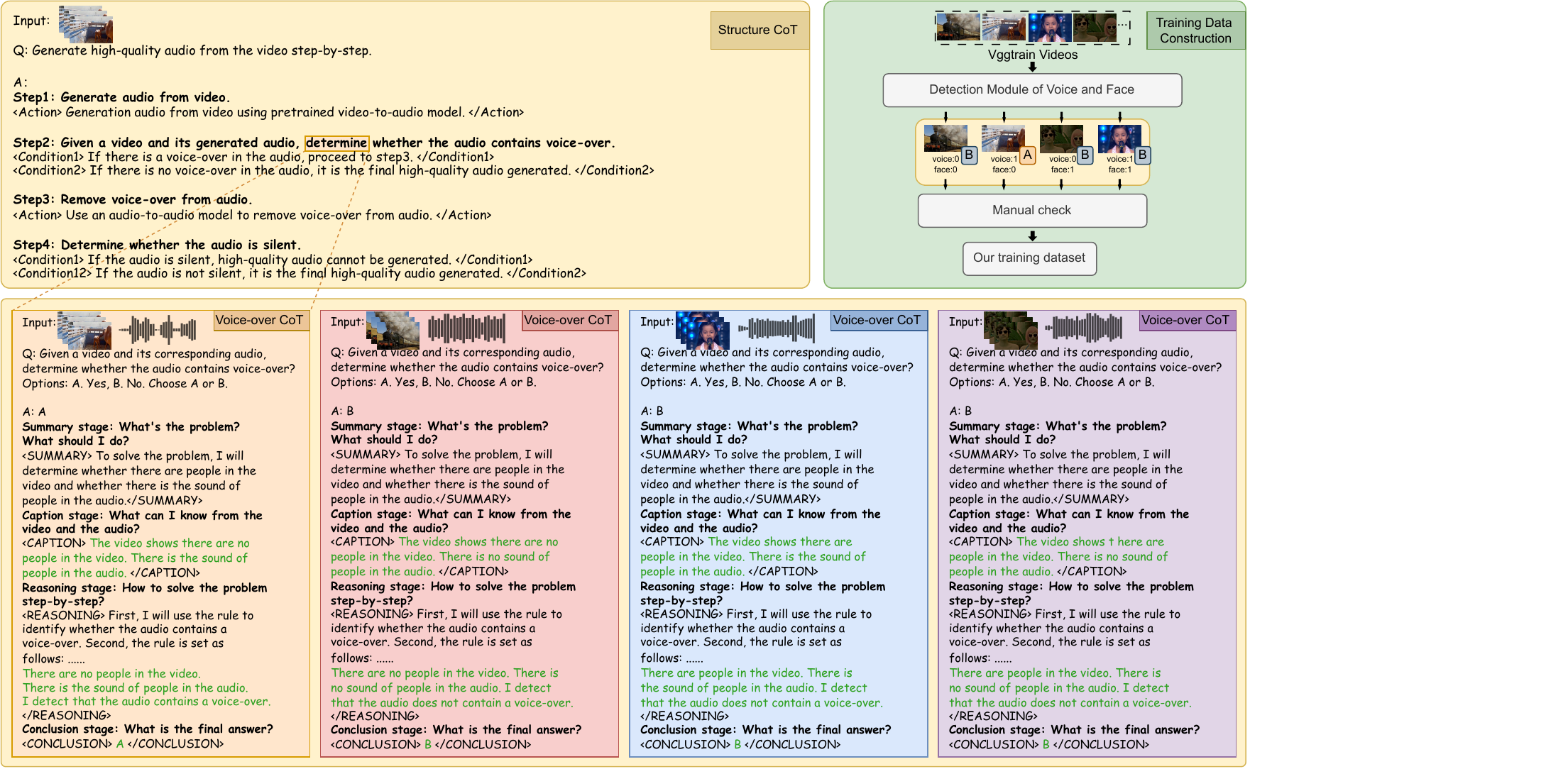}
    \caption{The process flow for generating the multiple CoT dataset involves utilizing multiple models and incorporating manual verification to ensure data quality.}
    \label{fig:fig4}
\end{figure*}

\input{tab/more}

\subsection{Implementation details} 
In the audio generation from video step, we adopt MMAudio\cite{cheng2024taming} as our baseline framework, a multimodal audio generation model that supports varying model sizes and sample rates. For the voice-over judgment step, we leverage VideoLLaMA2~\cite{cheng2024VideoLLaMA}, a large multimodal model with strong video understanding capabilities, supporting both audio and video inputs, as our MLLM baseline. In the voice-over removal from audio step, we use BS-Roformer~\cite{lu2024music} as the baseline, which demonstrates strong performance in human voice separation.

We use the constructed $18k$ $\text{CoT}_{\text{detail}}$ data and $1.8k$ $\text{CoT}_{\text{structure}}$ to fine-tune the VideoLLaMA2 model to support multiple related tasks. We fine-tune its audio-video joint stage, The video encoder remains frozen while we optimize the audio/video projector and the audio encoder, alongside the unfrozen LLM. The training is carried out with a batch size of $128$ and a total of $1$ epoch on $8$ Nvidia A800 GPUs. We use the AdamW optimizer with a learning rate of $2e-5$.

\subsection{Evaluation Metrics} 
We evaluate the generated audio from four perspectives: distribution matching, audio quality, semantic alignment, and temporal alignment. Following prior work~\cite{iashin2021taming, wang2024frieren}, we adopt Fréchet Distance (FD) and Kullback–Leibler (KL) divergence as metrics. For FD, we extract audio embeddings using PaSST~\cite{koutini2021efficient} ($FD_{PaSST}$), PANNs~\cite{kong2020panns} ($FD_{PANNs}$), and VGGish~\cite{gemmeke2017audio} ($FD_{VGG}$). For KL distance, we use PANNs ($KL_{PANNs}$) and PaSST ($KL_{PaSST}$) as classifiers, following~\cite{liu2023audioldm}. Similarly, Inception Score (IS)~\cite{salimans2016improved}, IB-score, DeSync are applied with the same setting as the state-of-the-art models ~\cite{wang2024frieren,girdhar2023imagebind,viertola2024temporally,iashin2024synchformer}. 

\input{tab/tab-main-results}

\begin{figure*}[t]
    \centering    \includegraphics[width=1\textwidth]{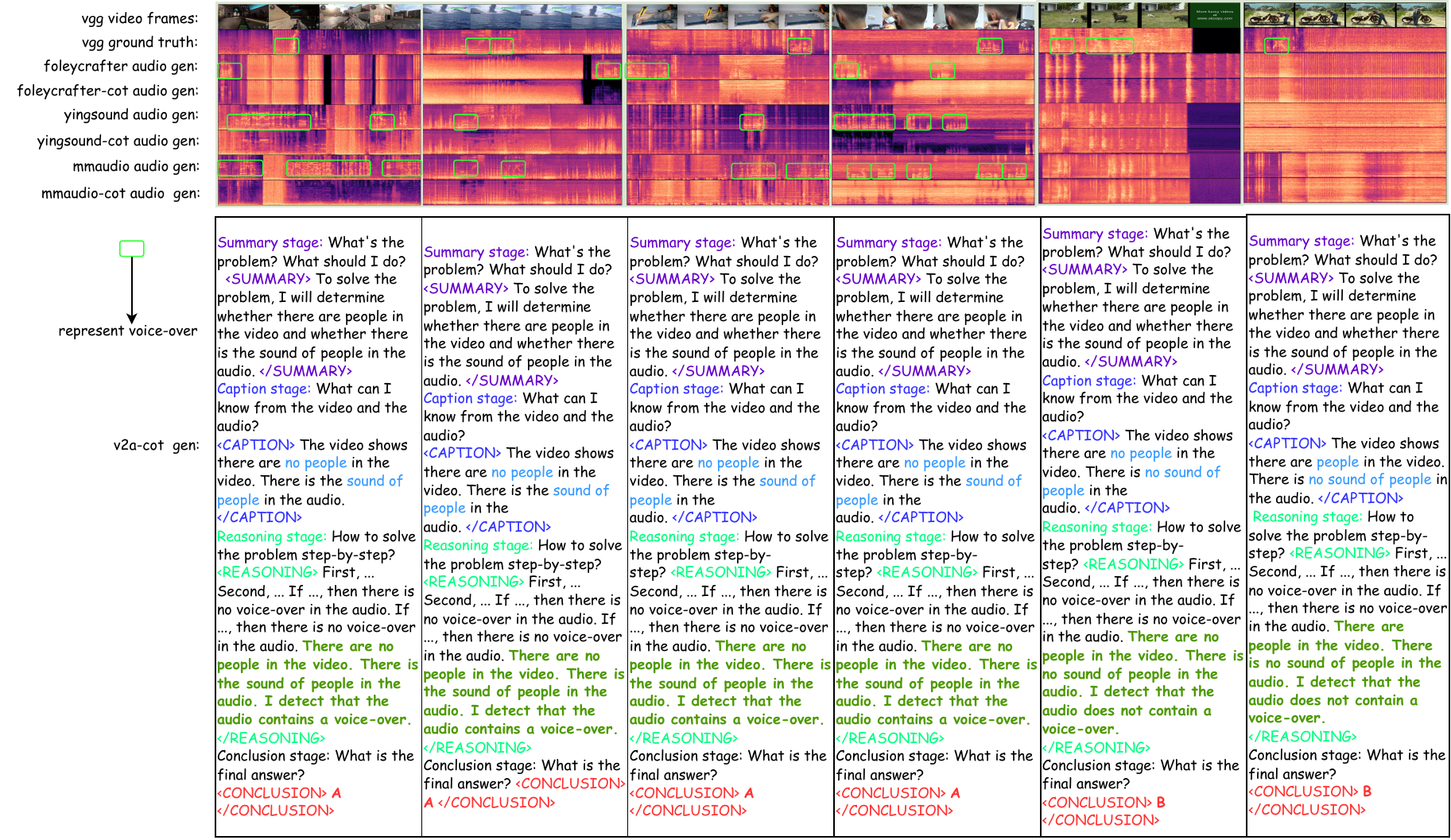}
    \caption{V2A-CoT results of our method.From top to bottom, the images are: the video frames from the VGGSound-test dataset, the mel-spectrogram of the ground truth audio, mel-spectrogram of the coarse audio generated by different models, mel-spectrogram of the final fine-grained audio generated by different models, and the CoT output of the voice-over detection.}
    \label{fig:fig5}
\end{figure*}

\begin{table}[ht]
\centering

\resizebox{\linewidth}{!}{
    \begin{tabular}{l|cccccc}
    \toprule
      Method & QA Ratio & CoT Ratio & QA Num & CoT Num & Total   \\ \midrule
    MMAudio-S-44k~\cite{cheng2024taming} & 40.46\%  & 58.30\%  & 1072  & 1455  & 1525 \\
    MMAudio-M-44k~\cite{cheng2024taming} & 40.61\%  & 57.22\%  & 1322  & 1726  & 1820 \\
    MMAudio-L-44k~\cite{cheng2024taming} & 40.78\%  & 55.62\%  & 1304  & 1675  & 1772  \\                  
    YingSound~\cite{chen2024yingsound} & 39.74\%   & 55.85\%    & 1459   & 1960  & 2020  \\
    FoleyCrafter~\cite{zhang2024foleycrafterbringsilentvideos} & 37.57\%   & 52.98\%    & 1220 & 1647 & 1708 \\ \bottomrule
    \end{tabular}
}
\caption{Results of Multimodal Large Model for Voice-over Detection. QA Ratio represents the energy ratio of human voices in the voice-over detected by QA, while CoT Ratio represents the energy ratio of human voices in the voice-over detected by CoT. QA Num and CoT Num indicate the number of voice-over detected by QA and CoT, respectively. Total represents the overall number of detected voice-over.}
\label{table:table3}
\end{table}

\subsection{Experiment Setting Details} 
We evaluate our method through multiple experiments. First, we test on two settings, the original VGGSound test set, referred to as \textbf{\textit{Ori-Set}}, and \textbf{\textit{VO-Free}}(Voice-Over-Free), which consists of $1436$ videos with voice-over in the original VGGSound test set that are converted into voiceover-free videos and the remain original videos.

Secondly, we introduce several types of generated audio for evaluation. The first type is the audio generated directly by the V2A model in Step 1, referred to as \textbf{\textit{Direct}}. Additionally, we include the audio generated by the V2A model with a negative prompt, referred to as \textbf{\textit{Direct-neg}}, where the negative prompt is: "human voice"

The second type is the audio generated by our framework through Step 1, Step 2, and Step 3, referred to as \textbf{\textit{Ours-s3}}. Based on the voiceover detection model used in Step 2, we denote the results as \textbf{\textit{Ours-s3}}, respectively. 

The third type is the audio generated by our complete framework through Step 1, Step 2, Step 3, and Step 4, referred to as \textbf{\textit{Ours-s4}}. Depending on the post-processing strategy in Step 4, the results are further divided into three variants: \textbf{\textit{Ours-s4-rm}}: After silence detection, the bar segment is removed if silence is detected.
\textbf{\textit{Ours-s4-rep}}: After silence detection, the bar segment is replaced with the corresponding audio generated by Step 1. \textbf{\textit{Ours-s4-neg}}: After silence detection, the bar segment is replaced with audio generated using a negative prompt: "human voice". 


\subsection{Main Results} 
We conducted experiments using MMAudio with various model sizes, as well as YingSound~\cite{chen2024yingsound} and FoleyCrafter~\cite{zhang2024foleycrafterbringsilentvideos} as the V2A model. The results, shown in Table~\ref{table:more}, demonstrate that the proposed method outperforms the baseline across several key metrics. Specifically, the $FD_{PaSST}$ metric is reduced by up to 10.07\%, $FD_{PANNS}$ by up to 11.62\%, and $FD_{VGG}$ by up to 38.61\%. Additionally, the IS indicator improves by up to 4.95\%, the IB-score increases by up to 6.39\%, and the DeSync metric is reduced by up to 0.89\%.

\subsection{Ablation Studies} 
To investigate the impact of different strategies on voiceover removal, we conduct three ablation experiments to evaluate the performance of our method under various conditions.

\textbf{Impact of Negative Prompt on Voice-over Removal.} To investigate the effect of prompts on voice-over removal, we introduce a negative prompt "human voice" during inference generation. As shown in Table~\ref{table:mmaduio-large-44k}, the negative prompt effects the quality of generated audio. 



\textbf{Impact of Reasoning for Voice-over Detection.} To further enhance the evaluation of voice-over presence, we leverage a multimodal large model to judge whether voice-over exist in the generated audio. The model is employed in two ways: direct QA for binary voiceover detection, and CoT reasoning to infer the final answer. Table~\ref{table:table3} presents the experimental results, CoT-reasoning obtains improvements over $15\%$ ratio for a variety of state-of-the-art V2A models. 

\textbf{The Impact of Post-processing Strategies for Silent Audio and Steps of Reasoning.} After removing voice-over, many generated audios only contain sounds that tend to be silent. In this case, we explore three post-processing strategies: (1) directly removing silence, (2) replacing silent segments with the original audio before voiceover removal, and (3) using the audio generated with the negative prompt directly as the final result. Table 3 shows that the first strategy achieves the highest proportion of overall metric improvement compared to the other two methods $8.93\%$ for $FD_{PaSST}$, $6.36\%$ for $FD_{PANNs}$ and $2.06\%$ for $FD_{VGG}$, the improvement between baseline and ours is represented as green color, demonstrating effectiveness of the learned CoT reasoning in enhancing the final audio quality. Besides, improvement of $0.57\%$, $0.83\%$, $0.69\%$ and $0.82\%$ are obtained in $KL_{PaSST}$, $IS$, $IB-score$ and $DeSync$ which is in the comparison between \textbf{\textit{Ours-s3}} and \textbf{\textit{Ours-s4}}, it's represented as blue color.

%% file: tab/more.tex
\begin{table*}[]

\centering
\resizebox{\linewidth}{!}{


\begin{tabular}{lcccccccc}
\toprule
                                                  Method & \multicolumn{5}{c}{Distribution matching}                                                                                                                        & Audio quality                & Semantic align                  & Temporal align                \\ \cline{2-9} 
{\color[HTML]{333333} }                            & {\color[HTML]{333333} $FD_{PaSST}\downarrow$} & {\color[HTML]{333333} $FD_{PANNs}\downarrow$} & {\color[HTML]{333333} $FD_{VGG}\downarrow$} & {\color[HTML]{333333} $KL_{PANNs}\downarrow$} & {\color[HTML]{333333} $KL_{PaSST}\downarrow$} & {\color[HTML]{333333} $IS\uparrow$} & {\color[HTML]{333333} $IB\text{-}score\uparrow$} & {\color[HTML]{333333} $DeSync\downarrow$} \\ \midrule



\multicolumn{6}{l}{\textbf{MMAudio-S-44k~\cite{cheng2024taming} $\downarrow$}} \\
        \toprule
{\color[HTML]{333333} Direct \& Ori-Set} & 65.25 & 5.55 & 1.66 & 1.67 & 1.44 & 18.02 & 32.27 & 0.444 \\
{\color[HTML]{333333} Direct \& VO-Free} & {\color[HTML]{333333} } 65.47 & {\color[HTML]{333333} } 5.77      & {\color[HTML]{333333} } 1.03    & {\color[HTML]{333333} } 2.22  & {\color[HTML]{333333} } 1.82  & {\color[HTML]{333333} } 13.32  & {\color[HTML]{333333} } 31.16 & {\color[HTML]{333333} } 0.487      \\
{\color[HTML]{333333} Direct-neg \& Ori-Set}  & {\color[HTML]{333333} } 68.44 & {\color[HTML]{333333} } 6.48  & {\color[HTML]{333333} } 1.71 & {\color[HTML]{333333} } 2.27  & {\color[HTML]{333333} } 1.84 & {\color[HTML]{333333} } 13.74  & {\color[HTML]{333333} } 30.51 & {\color[HTML]{333333} } 0.505      \\
{\color[HTML]{333333} Our best \& VO-Free} & {\color[HTML]{333333} } \textbf{65.07}(\textcolor[HTML]{006400}{0.27\%}) & {\color[HTML]{333333} } 6.08 & {\color[HTML]{333333} } \textbf{1.02}(\textcolor[HTML]{006400}{38.61\%}) & {\color[HTML]{333333} } 2.20  & {\color[HTML]{333333} } 1.82  & {\color[HTML]{333333} } 13.39  & {\color[HTML]{333333} } 30.82 & {\color[HTML]{333333} } 0.496    \\ \midrule

\multicolumn{6}{l}{\textbf{MMAudio-M-44k~\cite{cheng2024taming} $\downarrow$}} \\
        \toprule
{\color[HTML]{333333} Direct \& Ori-Set} & 61.88 & 4.74 & 1.13 & 1.66 & 1.41 & 17.41 & 32.99 & 0.443 \\
{\color[HTML]{333333} Direct \& VO-Free}     & {\color[HTML]{333333} } 56.07 & {\color[HTML]{333333} } 4.57  & {\color[HTML]{333333} } 0.99  & {\color[HTML]{333333} } 2.15  & {\color[HTML]{333333} } 1.74       & {\color[HTML]{333333} } 13.91  & {\color[HTML]{333333} } 32.19  & {\color[HTML]{333333} }  0.479     \\
{\color[HTML]{333333} Direct-neg \& Ori-Set}  & {\color[HTML]{333333} }  60.21  & {\color[HTML]{333333} } 4.79   & {\color[HTML]{333333} } 1.66  & {\color[HTML]{333333} } 2.20  & {\color[HTML]{333333} }  1.76      & {\color[HTML]{333333} } 14.68  & {\color[HTML]{333333} } 32.13  & {\color[HTML]{333333} } 0.486       \\

Our best \& VO-Free  &  \textbf{55.65}(\textcolor[HTML]{006400}{10.07\%}) &  4.80   & \textbf{0.93} (\textcolor[HTML]{006400}{17.70\%}) & 2.15   & 1.77   &  13.82 &  31.44    &  0.495  \\ \midrule

\multicolumn{6}{l}{\textbf{MMAudio-L-44k~\cite{cheng2024taming} $\downarrow$}} \\
        \toprule
{\color[HTML]{333333} Direct \& Ori-Set} & 60.60 & 4.72 & 0.97 & 1.65 & 1.40 & 17.40 & 33.22 & 0.442 \\
{\color[HTML]{333333} Direct \& VO-Free}     &  56.29  &  4.29 &  1.03  &  2.13  &  1.72  &  14.54 &   32.74  &  0.475                            \\
{\color[HTML]{333333} Direct-neg \& Ori-Set}  &   59.50 &  4.62  &   1.75 & 2.19  & 1.76  &  15.42 & 32.36 &  0.490                             \\

Our best \& VO-Free & \textbf{55.19} (\textcolor[HTML]{006400}{8.93\%})& \textbf{4.42} (\textcolor[HTML]{006400}{6.36\%})& \textbf{0.95} (\textcolor[HTML]{006400}{2.06\%}) & 2.13 & 1.75 & 14.49 & 31.94 & 0.490 \\ \midrule

\multicolumn{6}{l}{\textbf{YingSound~\cite{chen2024yingsound} $\downarrow$}} \\
        \toprule
{\color[HTML]{333333} Direct \& Ori-Set}     &   69.37   &  6.28 &  0.78   &  1.70  & 1.41 &  14.02  & 27.75  &  0.956                           \\
{\color[HTML]{333333} Direct \& VO-Free}     &  68.78 &   5.33   & 0.70  &  1.74  &  1.45   &  14.63  &   27.75  &  0.956                            \\
{\color[HTML]{333333} Direct-neg \& Ori-Set}  &  77.86   &  7.37 & 0.75  &  2.20 &   1.83  &  12.48 &  27.15  &  0.991                            \\
{\color[HTML]{333333} Our best \& VO-Free} &  \textbf{68.95}(\textcolor[HTML]{006400}{0.60\%}) & \textbf{5.57}(\textcolor[HTML]{006400}{11.32\%})   &  \textbf{0.72}(\textcolor[HTML]{006400}{8.32\%})  & 1.73  & 1.45  &  \textbf{14.71}(\textcolor[HTML]{006400}{4.95\%})   &  27.56  &  0.962  \\ 
\midrule

\multicolumn{6}{l}{\textbf{FoleyCrafter~\cite{zhang2024foleycrafterbringsilentvideos} $\downarrow$}} \\
        \toprule
{\color[HTML]{333333} Direct \& Ori-Set}     & 140.09 & 19.67 & 2.51 & 2.30 & 2.23 & 15.58 & 25.68 & 1.225    \\
{\color[HTML]{333333} Direct \& VO-Free}     &   130.67 & 17.59 & 2.12 & 2.59 & 2.28 & 9.94 & 27.96 & 1.215   \\
{\color[HTML]{333333} Direct-neg \& Ori-Set}  & 181.45 & 21.17 & 3.17 & 2.73 & 2.43 & 10.48 & 27.34 & 1.223 \\
{\color[HTML]{333333} Our best \& VO-Free} &  \textbf{127.97}(\textcolor[HTML]{006400}{8.65\%}) & \textbf{17.39}(\textcolor[HTML]{006400}{11.62\%}) & \textbf{2.12}(\textcolor[HTML]{006400}{15.42\%}) & 2.57 & 2.29 & 9.96 & \textbf{27.43}(\textcolor[HTML]{006400}{6.39\%}) & \textbf{1.214}(\textcolor[HTML]{006400}{0.89\%})     \\ 
\bottomrule
\end{tabular}
}

\caption{Video-to-audio results on the VGGSound test set. The bold text highlights the superior performance of our proposed method compared to previous methods, while the green text in brackets represents the improvement rate of each index.}
\label{table:more}

\end{table*}

%% file: tab/tab-main-results.tex
\begin{table*}[]

\centering
\resizebox{\linewidth}{!}
{

\begin{tabular}{lcccccccc}
\hline
                                                   & \multicolumn{5}{c}{Distribution matching}                                                                                                                        & Audio quality                & Semantic align                  & Temporal align                \\ \cline{2-9} 

& $FD_{PaSST}\downarrow$ & $FD_{PANNs}\downarrow$ &  $FD_{VGG}\downarrow$ & $KL_{PANNs}\downarrow$ &  $KL_{PaSST}\downarrow$ &  $IS\uparrow$ &  $IB\text{-}score\uparrow$ &  $DeSync\downarrow$ \\ \hline
 Direct \& Ori-Set      & 60.60 & 4.72 & 0.97 & 1.65 & 1.40 & 17.40 & 33.22 & 0.442 \\
 Direct \& VO-Free    &  56.29  &  4.29 &  1.03  &  2.13  &  1.72  &  14.54 &   32.74  &  0.475                            \\
Direct-neg \& Ori-Set  &   59.50 &  4.62  &   1.75 & 2.19  & 1.76  &  15.42 & 32.36 &  0.490                             \\ \midrule

Ours-s3 \& VO-Free & \textbf{55.19} (\textcolor[HTML]{006400}{8.93\%})& \textbf{4.42} (\textcolor[HTML]{006400}{6.36\%})& \textbf{0.95} (\textcolor[HTML]{006400}{2.06\%}) & 2.13 & 1.75 & 14.49 & 31.94 & 0.490 \\

Ours-s4-rm \& VO-Free  &  55.75 & 4.49 & 1.00 & 2.12 & 1.73 & 14.70 & 32.25 & 0.484 \\
Ours-s4-rep \& VO-Free &  \textbf{55.66}(\textcolor[HTML]{006400}{8.15\%}) & \textbf{4.45}(\textcolor[HTML]{006400}{5.72\%}) & \textbf{0.97}(\textcolor[HTML]{006400}{0.00\%}) & 2.14 & \textbf{1.74}(\textcolor[HTML]{00008B}{0.57\%}) & \textbf{14.61}(\textcolor[HTML]{00008B}{0.83\%}) & \textbf{32.16}(\textcolor[HTML]{00008B}{0.69\%}) & \textbf{0.486}(\textcolor[HTML]{00008B}{0.82\%}) \\
Ours-s4-neg \& VO-Free &  55.66 & 4.44 & 0.99 & 2.13 & 1.74 & 14.65 & 32.17 & 0.487  \\ \hline
\end{tabular}

}
\caption{Ablation result on MMAudio-L-44k. The improvement between baseline and ours is represented as green color, demonstrating effectiveness of the learned CoT reasoning in enhancing the final audio quality, the improvement between Ours-s3 and Ours-s4 is represented as blue color.}
\label{table:mmaduio-large-44k}
\end{table*}

%% file: sec/5_discussion.tex
\section{Discussion}
We propose DeepSound, an end-to-end framework that enables audio generation from videos through initial step-by-step thinking, based on the internal CoT of MLLM, without requiring additional annotations. A corresponding multi-modal reasoning dataset is constructed to support the learning of initial thinking in audio generation. We are currently developing the next version, which incorporates an in-depth thinking mechanism within a single network architecture.